\documentclass[fleqn,usenatbib,useAMS]{mnras}

\usepackage{mathptmx}
\usepackage{txfonts}

\usepackage[T1]{fontenc}
\usepackage{ae,aecompl}

\DeclareRobustCommand{\VAN}[3]{#2}
\let\VANthebibliography\thebibliography
\def\thebibliography{\DeclareRobustCommand{\VAN}[3]{##3}\VANthebibliography}

\usepackage{graphicx}	
\usepackage{bm}
\usepackage{widetext}
\usepackage{subfig,epsfig,epsf,color}
\usepackage[utf8x]{inputenc}
\usepackage{appendix}
\usepackage{slashed}
\usepackage{feynmf}
\usepackage{array}
\usepackage{pdflscape}	
\usepackage{hyperref}

\newcommand{\bea}{\begin{eqnarray}}
\newcommand{\eea}{\end{eqnarray}}

\title[Vector dark boson mediator in quark stars]{Vector dark boson mediated feeble interaction between fermionic dark matter and strange quark matter in quark stars}

\author[Debashree Sen and Atanu Guha]{
Debashree Sen,$^{1}$\thanks{E-mail: debashree@vecc.gov.in}
Atanu Guha,$^{2}$\thanks{E-mail: atanu@cnu.ac.kr}
\\
$^{1}$Physics Group, Variable Energy Cyclotron Centre, 1/AF Bidhan Nagar, Kolkata
700064, India\\
$^{2}$Department of Physics, Chungnam National University, 99, Daehak-ro, Yuseong-gu, Daejeon-34134, South Korea}

\date{Accepted XXX. Received YYY; in original form ZZZ}

\pubyear{2022}

\begin{document}
\label{firstpage}
\pagerange{\pageref{firstpage}--\pageref{lastpage}}
\maketitle

\begin{abstract}

We study the structural properties like the gravitational mass, radius and tidal deformability of dark matter (DM) admixed strange quark stars (SQSs). For the purpose we consider the vector MIT Bag model to describe the strange quark matter (SQM) and investigate the possible presence of accreted DM in the SQSs consequently forming DM admixed SQSs. We introduce feeble interaction between SQM and the accreted fermionic DM via a vector dark boson mediator. Considering the present literature, in the context of possible presence of DM in SQSs, this work is the first to consider interaction between DM and SQM in the DM admixed SQSs. The mass of the DM fermion ($m_{\chi}$) and the vector mediator ($m_{\xi}$) and the coupling ($y_{\xi}$) between them are determined in accordance with the constraint from Bullet cluster and the present day relic abundance, respectively. We find that the presence of DM reduces both the mass and radius of the star compared to the no-DM case. The massive the DM fermion, the lower the values of maximum mass and radius of the DM admixed SQSs. For the chosen values of $m_{\chi}$ and corresponding values of $m_{\xi}$ and $y_{\xi}$, the computed structural properties of the DM admixed SQSs satisfy all the various present day astrophysical constraints.{{We obtain massive DM admixed SQSs configurations consistent with the GW190814 observational data. Hence the secondary compact object associated with this event may be a DM admixed SQS.}}


\end{abstract}




\maketitle



\section{Introduction}
\label{Intro}

 Several unknown and inconclusive facets of compact stars make them one of the most interesting objects in the Universe. One such inconclusive fact is their composition at such high density ($5-10$ times the nuclear density). So the dense matter composition and interactions at conditions relevant to compact stars are at present best understood by theoretical modeling of compact star matter. The corresponding interactions and the equation of state (EoS) are thus obtained based on speculative studies on theoretical modeling of compact star matter. Consequently, the presence of quark matter (QM) in compact stars is still experimentally unknown and a topic of current research. Theoretical speculations have predicted the possible existence of strange quark stars (SQSs) \cite{Olinto:1986je} based on the Bodmer-Witten conjecture which states that strange QM (SQM), being composed of $u$, $d$ and $s$ quarks, have a lower energy per baryon number than the pure nucleonic system \cite{Bodmer:1971we,Chin:1979yb,Witten:1984rs}. Ref. \cite{Farhi:1984qu} also ensured the stability of SQM at large baryon number and no external pressure with certain QCD parameters. Consequently, several theoretical works attempted the modeling of SQM in order to establish the possible existence of SQSs. The first and one of the most widely adopted model is the original MIT Bag model \cite{Chodos:1974je} which was later modified in \cite{Fraga:2001id, Alford:2004pf} as the non-ideal bag model. Further, repulsive interaction between the quarks was introduced via a parameter $\alpha_4$ \cite{Fraga:2001id, Alford:2004pf,Glendenning:1997wn,Weissenborn:2011qu}. The repulsive effect was also included by introducing vector meson as mediator (vBag model) \cite{Klahn:2015mfa, Cierniak:2019hhe,Franzon:2016urz,Wei:2018mxy,Lopes:2020btp,Kumar:2022byc}.
 
 The bag constant $B$ associated with the MIT Bag model represent the difference in energy density between the perturbative vacuum and the true vacuum. The bag constant $B$ is still not well known. Model dependent analysis with respect to GW170817 data constrained $B^{1/4}=$(134.1 - 141.4) MeV with low-spin prior and $B^{1/4}=$(126.1 - 141.4) MeV with high-spin prior for SQSs \cite{Zhou:2017pha} while for hybrid stars \cite{Nandi:2017rhy,Nandi:2020luz} estimated similar range of $B$. Ref. \cite{Aziz:2019rgf} also obtained the allowed range of $B^{1/4}=$(133.68 - 222.53) MeV for SQSs while \cite{Yang:2019rxn} constrained $B^{1/4}=$(141.3 - 150.9) MeV for SQSs with modified Tolman-Oppenheimer-Volkoff (TOV) formalism. It is also seen that the stability of SQS in terms of binding energy per baryon $\varepsilon/\rho_B$ is controlled by $B$ \cite{Farhi:1984qu,Torres:2012xv,Ferrer:2015vca}. Ref. \cite{Torres:2012xv} estimated the allowed range of $B$ with respect to the stability condition of SQSs demanding that $\varepsilon/\rho_B \leq$ 930 MeV, where $\rho_B$ is the baryon density. In $\beta$ equilibrated SQM the upper bound on $B$ is set by considering charge neutral 3 flavor SQM in presence of electrons while the lower bound is obtained with 2 flavor QM \cite{Torres:2012xv}. Recently, \cite{Lopes:2020btp} calculated the same for the vBag model and established the allowed range of $B$ for different vector couplings $G_V$. In the present work we adopt the same vBag model following \cite{Lopes:2020btp} and calculate the stability window of $B$ following the same criteria as $\varepsilon/\rho_B \leq m_n$. To calculate the EoS of SQM we consider an average value of $B$ between $B_{max}$ and $B_{min}$ for a particular value of $G_V$.
 
 Certain observational evidences like the rotation curves of the galaxies, observation of gravitational lensing, X-ray analysis of Bullet cluster \cite{Bertone:2004pz, Aghanim:2018eyx} support the existence of dark matter (DM) in the Universe and compact stars like SQSs, being highly gravitating objects, are capable of accreting DM onto the system forming DM admixed SQSs. The exact nature and properties and interaction of DM particle candidates are unknown. The most suitable DM particle candidates are the Weakly Interacting Massive Particles (WIMPs) whose direct detection is being attempted in various experiments like superCDMS \cite{Agnese:2018col}, XENON100 \cite{Aprile:2012nq}, XENON1T \cite{Aprile:2018dbl}, LUX \cite{Akerib:2012ak}, PANDAX-II \cite{Wang:2020coa}, DARKSIDE-50 \cite{Agnes:2018oej}, SENSEI \cite{Crisler:2018gci} and very recently the LUX-ZEPLIN (LZ) \cite{LZ:2022ufs} etc. However, the exclusion bounds prescribed by such direct detection experiments are dependent on the local DM density around the solar neighborhood which do not affect the density of DM in the NS/SQS environment. The Cosmic Microwave Background (CMB) anisotropy maps, obtained from the Wilkinson Microwave Anisotropy Probe (WMAP) data \cite{WMAP:2012fli}, furnishes the present day thermal relic abundances of DM to be $\sim \Omega h^2 \approx 0.12$ \cite{ParticleDataGroup:2018ovx,Bauer:2017qwy,Cannoni:2015wba}. Thus any feasible DM model has to successfully reproduce the observed non-baryonic relic density. This sets strong constraints on the model parameters. On being accreted to the SQS system, the DM particles suffer collisions with the quarks and thereby lose kinetic energy and eventually become gravitationally bound to the star. At some point of time the accretion stops and the DM particles attain thermal equilibrium among themselves due to the self interactions. This justifies the DM particle density to be almost constant as considered by \cite{Panotopoulos:2017idn,Guha:2021njn,Sen:2021wev} in case of DM admixed neutron stars (NSs). The accrecred DM particles therefore remain confined within a region of small radius inside the star. Several works have successfully explained the possible existence of DM admixed SQSs and DM admixed NSs in which the SQM or the NS matter do not interact with the DM \cite{Lopes:2018oao,Ellis:2018bkr,Li:2012ii,Tolos:2015qra, Deliyergiyev:2019vti,Rezaei:2016zje,Mukhopadhyay:2016dsg,Mukhopadhyay:2015xhs,Panotopoulos:2017pgv,Jimenez:2021nmr,Panotopoulos:2018ipq,Leung:2022wcf,Karkevandi:2021ygv,Lourenco:2021dvh,Gleason:2022eeg,Dengler:2021qcq,Karkevandi:2021ygv,Panotopoulos:2018joc,Panotopoulos:2017eig,Miao:2022rqj} and the two fluid approach is mostly adopted in such works. In this context it is worth mentioning that the interaction between DM and SQM in the DM admixed SQSs is not considered before in the existing literature to the best of our knowledge.
 
 The accreted DM may eventually lead to the collapse of the star into a black hole. To prevent this, the interaction between DM and the standard model constituents of the star matter must be extremely weak \cite{Zheng:2016ygg}. Therefore in the present work we invoke feeble interaction between SQM and fermionic DM $\chi$. As stated earlier that for the description of SQM, in the present work we consider the vbag model that introduces a vector mediator ($\omega$ meson) to ensure quark interaction. Motivated by this fact, in the present work we include the vector new physics mediator $\xi$ to study the effects of the feeble DM-SQM interaction on the DM admixed SQS properties. We choose the dark boson mediator $\xi$ in the dark sector to be of vector type in order to maintain similarity with the pure SQM that involves the $\omega$ meson which is a vector mediator. Moreover, the vector $\omega$ meson and the vector dark boson $\xi$ carry the same net spin. It is well known that the net spin of $\omega$ meson is 1 while that of the vector dark boson is also 1 \cite{Hambye:2019tjt,Arcadi:2020jqf,Gabrielli:2015hua}. However, one can also consider a scalar mediator from the dark sector. Several works \cite{Panotopoulos:2017idn, Bertoni:2013bsa, Nelson:2018xtr, Bhat:2019tnz, Lourenco:2022fmf, Quddus:2019ghy, Das:2018frc, Das:2020vng, Das:2021yny} have considered interaction between dark and hadronic matter via Higgs boson and have successfully computed the structural properties of DM admixed NSs. In our earlier works we introduced feeble interaction between DM fermion and $\beta$ equilibrated hadronic NS matter (described by the hadronic relativistic mean field model \cite{Sen:2020edi}) via a dark scalar mediator $\phi$ \cite{Sen:2021wev} while in \cite{Guha:2021njn} we included same dark vector mediator $\xi$ (as chosen in the present work) along with the scalar dark mediator $\phi$ in order to determine the structural properties of the DM admixed NSs. In the present work we consider only the vector dark mediator $\xi$ in order to be consistent with the SQM sector which involves only vector mediator and also to match the net spin of the $\omega$ and $\xi$ vector mediators. Similar to \cite{Guha:2021njn}, in the present work the mass of DM fermion $m_{\chi}$, mass of vector mediator $m_{\xi}$ and the coupling $y_\xi$ between them are determined in consistence with the self-interaction constraint from Bullet cluster \cite{Tulin:2013teo, Tulin:2017ara,Hambye:2019tjt} and from present day relic abundance \cite{Belanger:2013oya,Gondolo:1990dk,Guha:2018mli}. In the present work we study the effects of variation of these DM parameters $m_{\chi}$, $m_{\xi}$ and $y_\xi$ on the DM admixed SQSs in the light of various astrophysical constraints like the lower bound on maximum mass from PSR J0740+6620 \cite{Fonseca:2021wxt}, the constraint on tidal deformability of a 1.4 $M_{\odot}$ from GW170817 \cite{LIGOScientific:2018cki} and the NICER data for PSR J0030+0451 \cite{Riley:2019yda,Miller:2019cac}. These astrophysical constraints on the structural properties of compact stars help us to obtain realistic EoS of compact stars to certain extent and to understand the possible composition of matter at such conditions. The nature of the massive secondary compact object associated with the detection of GW190814 \cite{LIGOScientific:2020zkf} has remained inconclusive whether this object is a black hole (BH) or a NS. This is because any further information related to GW190814 like its electromagnetic counterpart or the tidal deformability of this compact object is still not obtained.

 The paper is organized as follows. In the next section \ref{Formalism} we depict the formalism of the vbag model for SQSs in the presence of DM. In the following section \ref{Results} we present our results along with relevant discussions. We finally conclude in the final section \ref{Conclusion} of the paper.


\section{Formalism}
\label{Formalism}

\subsection{vBag Model with Dark Matter}

 We adopt the vector MIT Bag (vBag) model with the $u$, $d$ and $s$ quarks in presence of electrons. The mass of $u$ and $d$ quarks is very small compared to that of the $s$ quark ($m_s$=95 MeV). The formalism of obtaining the EoS of such a system is well depicted in \cite{Lopes:2020btp,Kumar:2022byc}. In addition to SQM in the model, we consider the presence of accreted DM. The interaction between the quarks and fermionic DM $\chi$ is mediated by the vector dark boson $\xi$ of mass $m_{\xi}$. The complete modified Lagrangian of such a system is given as 
 
\bea
\mathcal{L} &=& \sum_f \bigg[\overline{\psi}_f \bigg(i \gamma_{\mu} \partial^{\mu} -g_{\xi}\gamma_{\mu}\xi^{\mu} - m_f \bigg) \psi_f - B \bigg] \Theta (\overline{\psi}_f \psi_f) \nonumber \\  &-& \sum_f g_{qqV} \bigg[\overline{\psi}_f \bigg(\gamma_{\mu}V^{\mu}\bigg) \psi_f \bigg] \Theta (\overline{\psi}_f \psi_f) + \frac{1}{2} m_V^2 V_{\mu} V^{\mu} - \frac{1}{4} V_{\mu\nu} V^{\mu\nu} \nonumber \\  &+& \overline{\psi}_l \bigg(i \gamma_{\mu} \partial^{\mu} - m_l\bigg) \psi_l - \frac{1}{4} F'_{\mu\nu} F'^{\mu\nu} + \frac{1}{2} m_{\xi}^2 \xi_{\mu} \xi^{\mu} \nonumber \\ &+& \overline{\chi}\bigg[i \gamma_{\mu} \partial^{\mu} -y_{\xi} \gamma_{\mu} \xi^{\mu} \bigg]\chi 
\protect\label{Eq:Lagrangian}
\eea 

 where, $f=$ $u$, $d$ and $s$ and the lepton $l$=e is the electrons. $B$ is the Bag constant and the Heaviside function $\Theta$=1 inside the bag. Here the quark interaction is mediated by the repulsive vector $\omega$ meson channel and thus $m_V=$783 MeV with $g_{qqV}$ as the coupling strength. For simplicity, in the present work we consider universal coupling scheme i.e., $g_{uuV}=g_{ddV}=g_{ssV}=g_{qqV}$. The scaled coupling is defined as $G_V=(g_{qqV}/m_V)^2$. So taking $G_V=$0 reduces to the original form of the MIT Bag model without interactions. Refs. \cite{Lopes:2020btp,Kumar:2022byc} also introduced the self-interaction of the vector $\omega$ field via its quartic contribution in terms of a parameter $b_4$ that regulates the increment/decrement of the vacuum expectation value ($V_0$) of the $\omega$ field. This correction term also mimics the Dirac sea contribution of the quarks. However, in the present work we do not consider this self-interaction of the $\omega$ field since we intend to study the exclusive effects of DM and the DM parameters on the structural properties of the DM admixed SQSs.
We consider the values of $G_V$=0.3, 0.5 and 0.7. As discussed in the Introduction section \ref{Intro}, in order to calculate the stability window for $B$, we adopt the same procedure as \cite{Torres:2012xv,Lopes:2020btp,Ferrer:2015vca} based on the criteria as $\varepsilon/\rho_B \leq m_n$ for the upper bound of $B$ with 3 flavor SQM while the lower bound is obtained with the 2 flavor QM. {{Here we consider, $m_n$=939 MeV to obtain the stability window for $B$.}} We present the calculated allowed range of $B$ for the considered values of $G_V$ in table \ref{table-stability}. Our calculated allowed range of $B$ is slightly different from that of \cite{Torres:2012xv,Lopes:2020btp} since unlike them we have considered $m_n$=939 MeV. Consistent with \cite{Lopes:2020btp}, we find that the value of both $B_{min}$ and $B_{max}$ decrease with increase of $G_V$. To calculate the EoS given using Eqs. (\ref{e}) and (\ref{P}), we consider an average value of $B$ between $B_{max}$ and $B_{min}$ for a particular value of $G_V$.

\begin{table}
\caption{Stability window obtained for the vector MIT Bag model with $X_V=1$ and $m_n$=939 MeV.}
{{
\setlength{\tabcolsep}{25.5pt}
\begin{center}
\begin{tabular}{ c c c c c c c c}
\hline
\hline
$G_V$ & $B^{1/4}_{min}$ & $B^{1/4}_{max}$ \\
      &(MeV)  &(MeV)  \\
\hline
\hline
0.3 & 138 & 148 \\
0.5 & 134 & 143 \\
0.7 & 131 & 139 \\  
\hline
\hline
\end{tabular}
\end{center}
}}
\protect\label{table-stability}
\end{table}  
 
\begin{figure}
\centering
\includegraphics[width=0.5\textwidth]{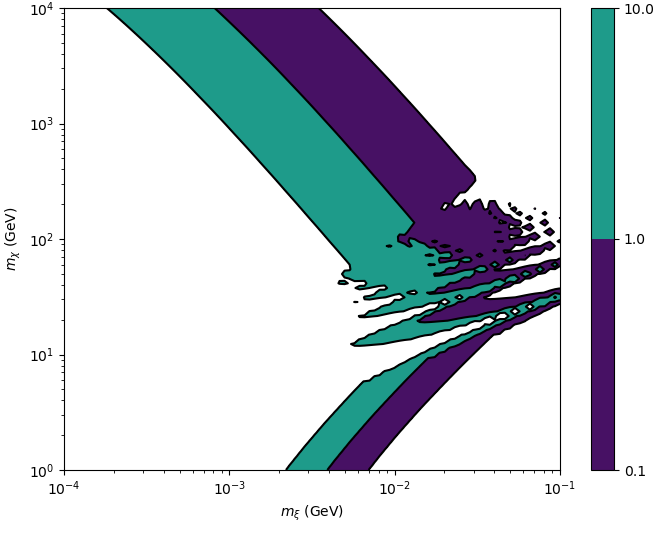}
\caption{Combination of $m_\chi$ and $m_\xi$ satisfying the self-interaction constraint from bullet cluster~\protect\cite{Randall:2007ph, Tulin:2013teo}. The color coding denotes the values of DM self-interaction transfer crossection $\sigma_T/m_\chi = (0.1\rm{-}1)~\rm{cm^2/gm}$ (purple) and $(1\rm{-}10)~\rm{cm^2/gm}$ (sea-green).}
\label{bullet_cluster}
\end{figure}

 The terms $\overline{\psi}_f(g_\xi \gamma_\mu \xi^\mu)\psi_f$ of Eq. (\ref{Eq:Lagrangian}) indicate the interaction of the the quark fields $\psi_f$ with the vector new physics mediator $\xi$ from dark sector with a very feeble coupling strength $g_\xi\sim 10^{-4}$ as we assumed in our previous works \cite{Guha:2021njn,Sen:2021wev}. We have checked that this coupling, being extremely small, the change in its order below $10^{-4}$ do not bring any significant change to the EoS and structure of DM admixed SQSs. The interaction between the fermionic DM $\chi$ and $\xi$ is depicted in the last line of Eq. (\ref{Eq:Lagrangian}). The corresponding coupling between the two is denoted by $y_{\xi}$. The terms $\frac{1}{4} V_{\mu\nu} V^{\mu\nu}$ and $\frac{1}{4} F'_{\mu\nu} F'^{\mu\nu}$ in Eq. (\ref{Eq:Lagrangian}) denote the kinetic terms of the vectors fields $\omega$ and $\xi$, respectively. In the present work the higher order self-interaction terms of the dark mediator $\xi$ has been neglected and has been chosen up to second order for simplicity. It is seen that the inclusion of the higher order terms do not contribute much and do not bring any substantial change to the results presented in the present work. The mass of the fermionic DM ($m_{\chi}$) and the corresponding value of $m_{\xi}$ is chosen in consistence with the self-interaction constraint from Bullet cluster \cite{Tulin:2013teo, Tulin:2017ara,Hambye:2019tjt} while the corresponding value of $y_{\xi}$ is chosen in accordance to the present day relic abundance \cite{Belanger:2013oya,Gondolo:1990dk,Guha:2018mli}. In our earlier work \cite{Guha:2021njn} we have already discussed in details the calculations of $m_{\chi}$, $m_{\xi}$ and $y_{\xi}$ consistent with the mentioned constraints. However, for the sake of completeness, we discuss the DM parameters and their calculations. Bullet cluster observational data suggests an estimate of the self-interaction of the DM particles \cite{Randall:2007ph, Bradac:2006er}. The self-scattering transfer cross-section of DM fermions is typically in the range $\sigma_T/m_\chi \approx (0.1\rm{-}10)~\rm{cm^2/gm}$ \cite{Randall:2007ph,Bradac:2006er,Dawson:2011kf,Dave:2000ar,Vogelsberger:2012ku,Kahlhoefer:2015vua}. For the DM fermions of mass $m_\chi$, self-scattered through the light vector mediator of mass $m_\xi$, the bullet cluster data estimates for the transfer cross-section as $\sigma_T/m_\chi \leq 1.25~\rm{cm^2/gm}$ \cite{Randall:2007ph, Robertson:2016xjh}. The DM parameters involved in the present work viz. the mass ($m_\chi$) of the DM fermions $\chi$ and the mass ($m_\xi$) of the light mediator $\xi$ satisfy the self-interaction constraints from bullet cluster \cite{Tulin:2013teo, Tulin:2017ara,Hambye:2019tjt} as seen from Fig. \ref{bullet_cluster}. The coupling constant ($y_\xi$) for the interaction between the DM fermion $\chi$ and the light vector mediators from the hidden sector $\xi$ is determined by satisfying the present day thermal relic abundances of DM. For detailed calculations one can refer to \cite{Belanger:2013oya, Gondolo:1990dk, Guha:2018mli}. The chosen parameter sets of the dark sector are listed in table \ref{table_DM}. 

\begin{table}
\caption{Chosen values of self interacting DM $m_\chi$  and corresponding values of $m_\xi$ from the constraints obtained from Bullet cluster. $y_\xi$ have been fixed from observed relic abundance.}
{{
\setlength{\tabcolsep}{25.5pt}
\begin{center}
\begin{tabular}{ c c c }
\hline
\hline
 $m_{\chi}$ & $m_{\xi}$ & $y_{\xi}$ \\
 (GeV) &(MeV) &  \\
\hline
\hline 
$25$ & $50$ & $0.27$ \\
$50$ & $60$ & $0.32$ \\ 
$75$ & $20$ & $0.40$ \\
$100$ & $10$ & $0.46$ \\
$150$ & $6$ & $0.50$ \\
$200$ & $4$ & $0.52$ \\
\hline\hline
\end{tabular}
\end{center}
}}
\protect\label{table_DM}
\end{table}

 Applying mean field approximation, the equation of motion of the vector field in terms of $V_0$ reads

\bea
V_0=\frac{g_{qqV}}{m_V^2}\rho
\eea 

where, the total quark density is given as

\begin{eqnarray}
\rho=<\psi_f^\dagger \psi_f>=(\rho_u + \rho_d + \rho_s)=\frac{\gamma_f}{6 \pi^2} \sum_f {{k_F}_f}^3
\label{rho}
\end{eqnarray} 

 Here ${k_F}_f$ is the Fermi momenta of the quarks and $\gamma_f=$6 for quarks. 

 The vacuum expectation value ($\xi_0$) of the vector dark boson is
 
\begin{eqnarray}
\xi_0=\frac{g_{\xi}\rho + y_{\xi}\rho_{\chi}}{m_{\xi}^2}
\end{eqnarray} 

where, the DM density is 

\begin{eqnarray}
\rho_{\chi}=\frac{\gamma_f}{6 \pi^2}{{{k_F}}_\chi}^3
\label{rho_chi}
\end{eqnarray}

 Based on the discussion given in \cite{Panotopoulos:2017idn,Guha:2021njn}, for the present work a constant number density of the fermionic DM has been assumed throughout the radius profile of the star. This estimate of constant number density ($\rho_{\chi}$) of fermionic DM is $\sim 1000$ times smaller than the average baryon number density of the SQM. In absence of any proper experimental or observational evidence regarding the presence and distribution of DM in compact stars, many recent works \cite{Bhat:2019tnz,Sen:2021wev,Guha:2021njn, Quddus:2019ghy, Das:2018frc, Das:2020vng, Das:2021yny} have also considered the same assumption of constant number density of DM throughout the density profile of NSs. Going with this assumption, the Fermi momentum of the DM fermions ${k_F}_{\chi}$ also turns out to be constant. Therefore in the present work we fix ${k_F}_{\chi}$=0.03 GeV.

 The quark chemical potential in presence of DM is modified as
 
\begin{eqnarray}
\mu_f=\sqrt{k_f^2 + m_f^2} + g_{qqV}V_0 + g_\xi \xi_0
\end{eqnarray}

The chemical equilibrium conditions and the charge neutrality condition for the quarks are additionally imposed.

The complete EoS is given as follows. The energy density is

\begin{eqnarray}
\varepsilon=\frac{1}{2}\frac{g_{qqV}^2}{m_V^2}\rho^2 + \frac{1}{2} \frac{\Big(g_{\xi}\rho + y_{\xi}\rho_{\chi}\Big)^2}{m_{\xi}^2} + \frac{\gamma_f}{2\pi^2}\sum_f \int_0^{{k_F}_f} \sqrt{k_f^2 + m_f^2}~ k_f^2~ dk_f \nonumber \\ + \frac{\gamma_l}{2\pi^2} \int_0^{{{k_F}}_l} \sqrt{k_l^2 + m_l^2}~ k_l^2~ dk_l + \frac{\gamma_\chi}{2\pi^2} \int_0^{{{k_F}}_\chi} \sqrt{k_\chi^2 + m_\chi^2}~ k_\chi^2~ dk_\chi + B
\label{e}
\end{eqnarray}

while the pressure is

\begin{eqnarray}
P=\frac{1}{2}\frac{g_{qqV}^2}{m_V^2}\rho^2 + \frac{1}{2} \frac{\Big(g_{\xi}\rho + y_{\xi}\rho_{\chi}\Big)^2}{m_{\xi}^2} + \frac{\gamma_f}{6\pi^2}\sum_f \int_0^{{k_F}_f} \frac{k_f^4~ dk_f}{\sqrt{k_f^2+m_f^2}} \nonumber \\ +  \frac{\gamma_l}{6\pi^2} \int_0^{{{k_F}}_l} \frac{{{k_l}}^4~ dk_l}{\sqrt{k_l^2+m_l^2}} + \frac{\gamma_l}{6\pi^2} \int_0^{{{k_F}}_\chi} \frac{{{k_\chi}}^4~ dk_\chi}{\sqrt{k_\chi^2+m_\chi^2}} - B
\label{P}
\end{eqnarray}

 With the obtained EoS we proceed to calculate the structural properties of the SQSs.

\subsection{Structural Properties of Quark Stars}
\label{structure}  
 
 With the obtained EoS, the structural properties like the gravitational mass ($M$) and the radius ($R$) of the DM admixed SQSs are computed by integrating the following Tolman-Oppenheimer-Volkoff (TOV) equations \cite{Tolman:1939jz,Oppenheimer:1939ne} based on the hydrostatic equilibrium between gravity and the internal pressure of the star.

\begin{eqnarray}
\frac{dP}{dr}=-\frac{G}{r}\frac{\left(\varepsilon+P\right)
\left(M+4\pi r^3 P\right)}{(r-2 GM)},
\label{tov}
\end{eqnarray}

\begin{eqnarray}
\frac{dM}{dr}= 4\pi r^2 \varepsilon,
\label{tov2}
\end{eqnarray}  

 The deformation of the metric $h_{\alpha \beta}$ in Regge-Wheeler gauge is given as \cite{Hinderer:2007mb,Hinderer:2009ca}
 
\begin{eqnarray} 
h_{\alpha\beta}=diag\left[e^{-\nu(r)}H_0,e^{\lambda(r)}H_2,r^2K(r),r^2\sin^2\theta K(r)\right]
Y_{2m}(\theta,\phi)
\label{h}
\end{eqnarray} 
 
 The tidal Love number $k_2$ is obtained in terms of the compactness ($C=M/R$) and a quantity $y$ which in case of QSs is defined as \cite{Hinderer:2009ca,Kumar:2022byc}
 
\begin{eqnarray}
y=\frac{RH'(R)}{H(R)} - \frac{4\pi R^3 \varepsilon_s}{M(R)}
\label{y}
\end{eqnarray}

where, $\varepsilon_s$ is the energy density at the surface of the QS. The tidal deformability parameter $\lambda$ in terms of $k_2$ is given as

\begin{eqnarray} 
\lambda=\frac{2}{3} k_2 R^5
\end{eqnarray}

 The dimensionless tidal deformability $\Lambda$ is then calculated as a function of Love number, gravitational mass and radius \cite{Hinderer:2007mb,Hinderer:2009ca} as
 
\begin{eqnarray} 
\Lambda=\frac{2}{3} k_2 (R/M)^5
\label{Lam}
\end{eqnarray}


\section{Results}
\label{Results}

 We compute the EoS of SQS both in presence and absence of DM. For the purpose we consider the values of $G_V$=0.3, 0.5 and 0.7. In the present work, for each value of $G_V$ we calculate the EoS with a corresponding value of $B$ which is the average of $B_{max}$ and $B_{min}$ presented in table \ref{table-stability}. The DM admixed SQM EoS is computed for three values of $m_{\chi}$=50, 75 and 100 GeV. For each value of $m_{\chi}$, $m_{\xi}$ and $y_{\xi}$ are calculated following \cite{Guha:2021njn} in order to be consistent with the constraints from Bullet cluster and present day relic density bound. With the obtained EoS for SQM and DM admixed SQM, we obtain the structural properties like the mass $M$, radius $R$ and tidal deformability $\Lambda$ of both SQSs and DM admixed SQSs following the formalism mentioned in section \ref{structure}.

\begin{figure}
\centering
\subfloat[]{\includegraphics[width=0.5\textwidth]{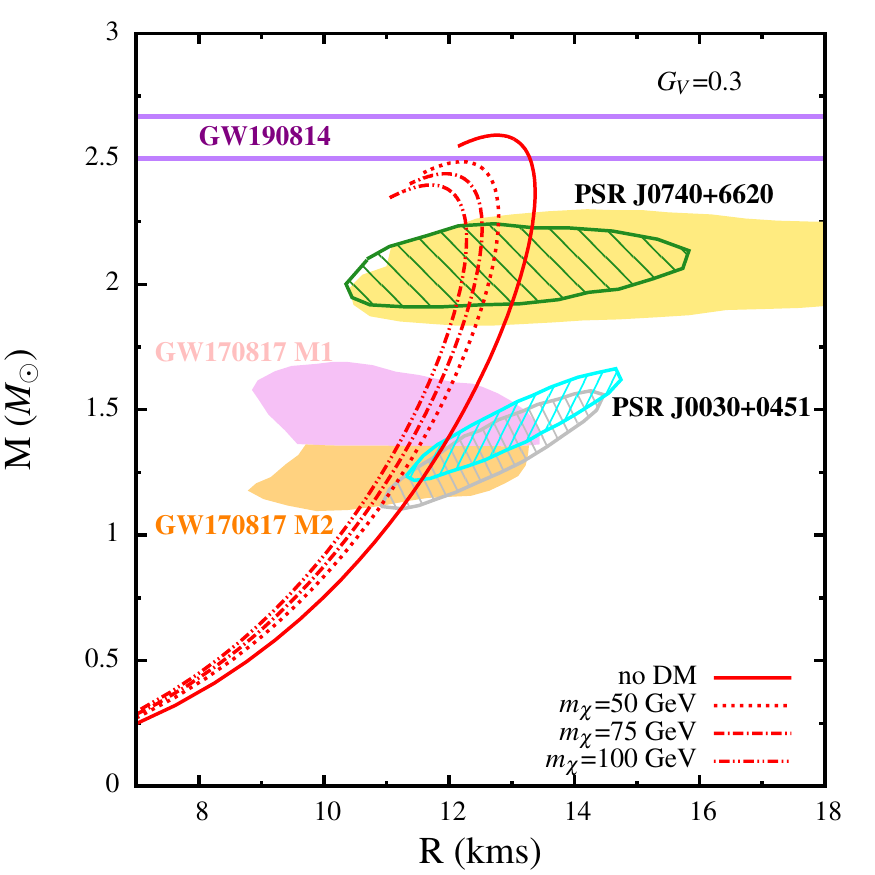}\protect\label{mr0p3}}
\vfill
\subfloat[]{\includegraphics[width=0.5\textwidth]{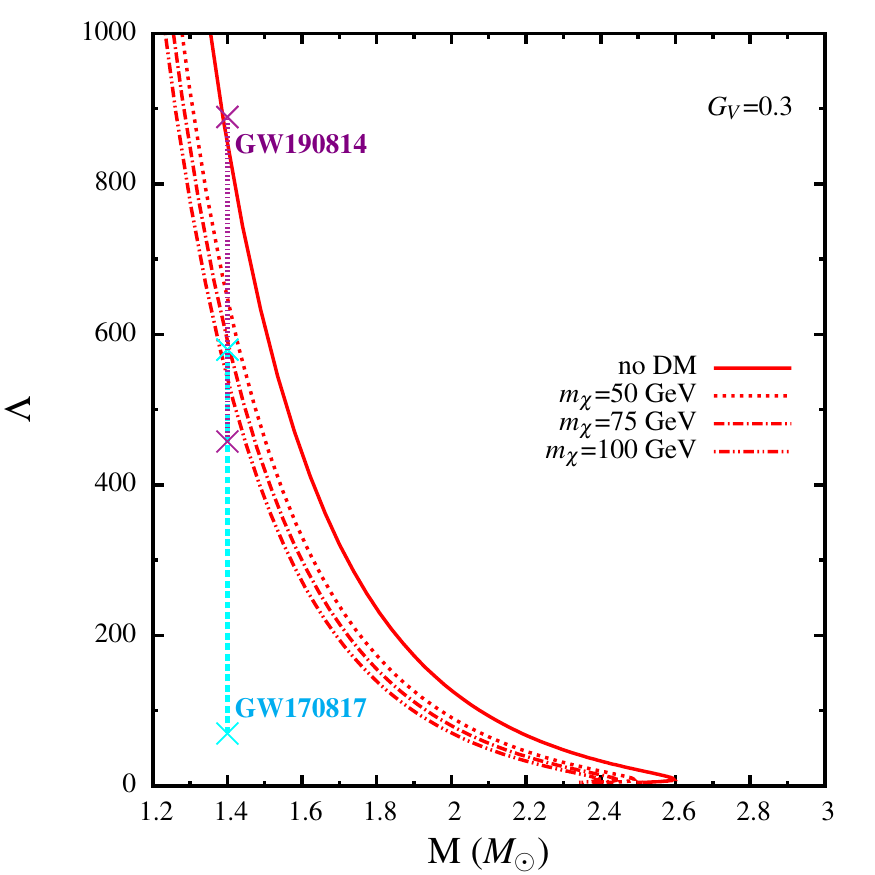}\protect\label{mLam0p3}}
\caption{\it (a) Mass-radius relationship of dark matter admixed quark stars for different $G_V$=0.3. Observational limits imposed from the most massive pulsar PSR J0740+6620 \protect\cite{Fonseca:2021wxt,Miller:2021qha,Riley:2021pdl} are also indicated. Mass of secondary component of GW190814 \protect\cite{LIGOScientific:2020zkf}) is also shown. The constraints on $M-R$ plane prescribed from GW170817 \protect\cite{LIGOScientific:2018cki}) and NICER experiment for PSR J0030+0451 \protect\cite{Riley:2019yda,Miller:2019cac} are also compared. (b) Corresponding variation of tidal deformability with respect to mass. Constraint on $\Lambda_{1.4}$ from GW170817 \protect\cite{LIGOScientific:2018cki} and GW190814 \protect\cite{LIGOScientific:2020zkf} observations are also shown.}
\label{MRLam0p3}
\end{figure}

 In Fig. \ref{MRLam0p3} we show our results of mass and radius (Fig. \ref{mr0p3}) and tidal deformability (Fig. \ref{mLam0p3}) with a lower value of $G_V$=0.3 and corresponding value of $B$ for both SQSs and DM admixed SQSs. From Fig. \ref{mr0p3} we find that the maximum mass $M_{max}$ of the star decreases as more massive DM fermion is considered. Compared to the no-DM (SQS) case, the constraint on mass from the secondary component of GW190814 \cite{LIGOScientific:2020zkf} is hardly satisfied in case of DM admixed SQSs. However, it is still debatable in literature whether this object is a compact star or a black hole. For both SQSs and DM admixed SQSs the $M-R$ constraints from GW170817 \cite{LIGOScientific:2018cki} and PSR J0740+6620 \cite{Fonseca:2021wxt,Miller:2021qha,Riley:2021pdl} are well satisfied. However, the constraint on the $M-R$ plane from NICER data for PSR J0030+0451 \cite{Riley:2019yda,Miller:2019cac}, though satisfied by SQS, is hardly satisfied by the DM admixed SQSs configurations. In Fig. \ref{mLam0p3} we find that for the low value of $G_V$(=0.3) the constraints on $\Lambda_{1.4}$ from both GW170817 and GW190814 are satisfied with $m_{\chi}$=75 and 100 GeV. For $m_{\chi}$=50 GeV and the no-DM only the constraint from GW190814 is satisfied. For the no-DM case our result is consistent with \cite{Kumar:2022byc}.
 
\begin{figure}
\centering
\subfloat[]{\includegraphics[width=0.5\textwidth]{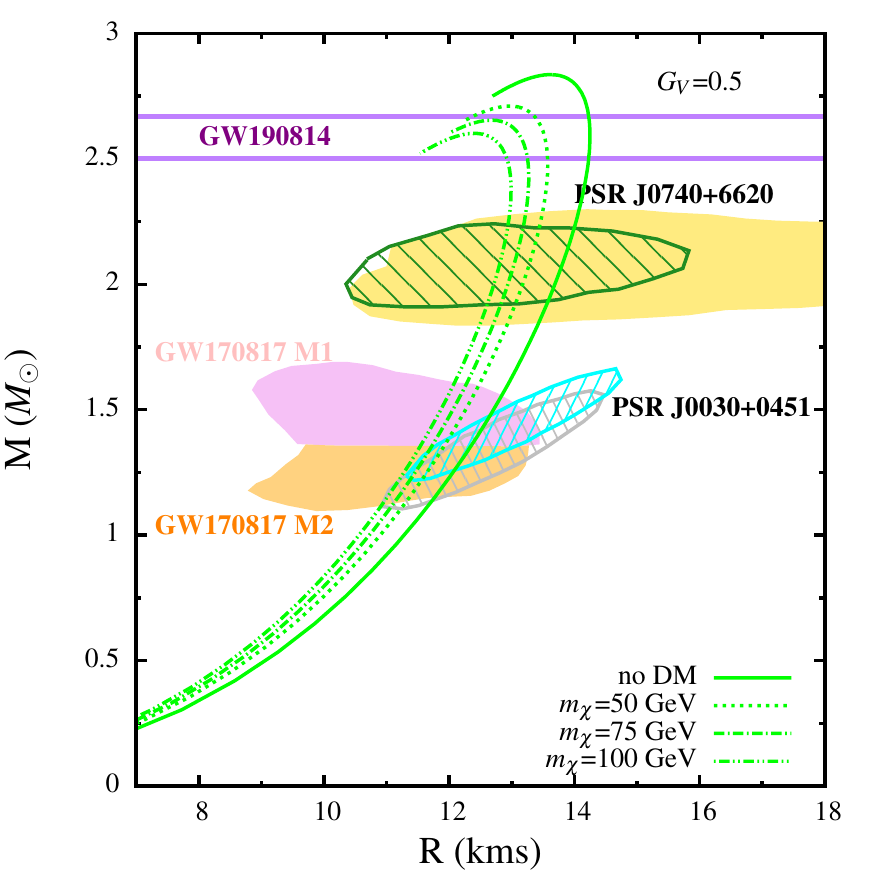}\protect\label{mr0p5}}
\vfill
\subfloat[]{\includegraphics[width=0.5\textwidth]{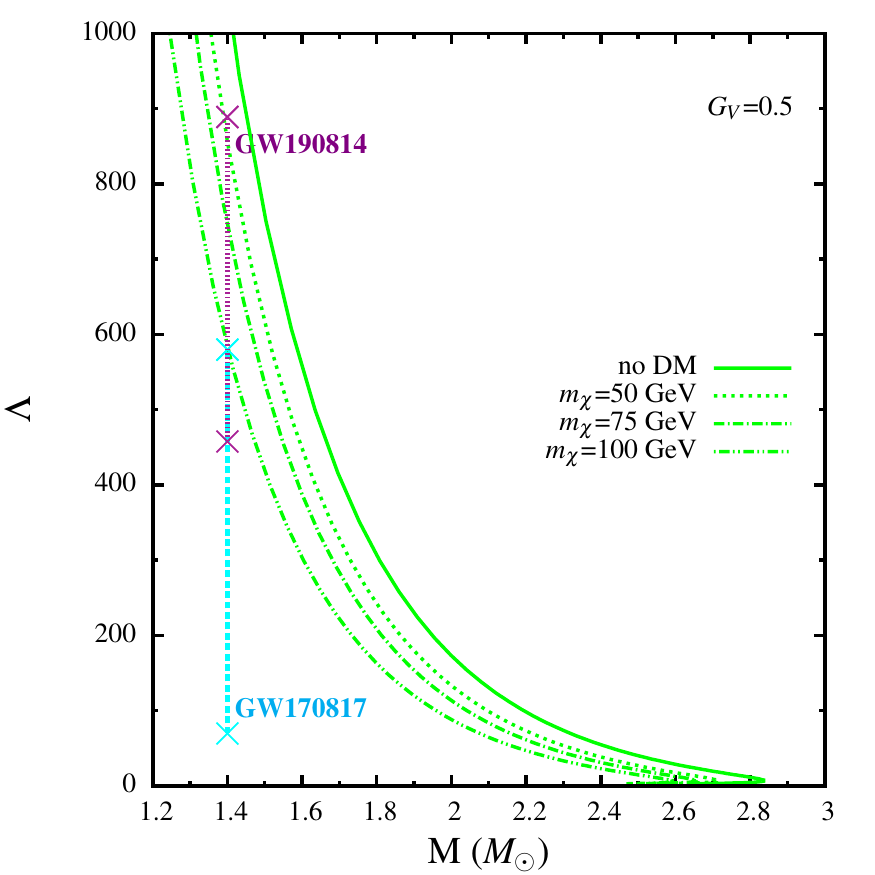}\protect\label{mLam0p5}}
\caption{\it Same as Fig. \ref{MRLam0p3} but for $G_V$=0.5.}
\label{MRLam0p5}
\end{figure}

 We next investigate the effect on the structural properties of SQSs and DM admixed SQSs in Fig. \ref{MRLam0p5} with an intermediate value of $G_V$=0.5 and the corresponding value of $B$. For DM admixed SQSs the same values are $m_{\chi}$ are considered same as that in case of Fig. \ref{MRLam0p3}. Comparing Figs. \ref{mr0p3} and \ref{mr0p5} we find that both $M_{max}$ and the corresponding radius of SQSs increase with increasing values of $G_V$. This is also consistent with \cite{Lopes:2020btp,Kumar:2022byc}. This fact is also true for the DM admixed SQSs for any particular value of $m_{\chi}$. For DM admixed SQSs the various constraints on the $M-R$ plane are better satisfied with a higher value of $G_V$. From Fig. \ref{mr0p5} we find that the constraint from NICER data for PSR J0030+0451, which was hardly satisfied with $G_V$=0.3, is now well satisfied for $G_V$=0.5 with all the three chosen values of $m_{\chi}$. We also note that unlike the DM admixed SQSs obtained with $G_V$=0.3, constraint on mass from the secondary component of GW190814 is well satisfied by that with $G_V$=0.5. Similar to the case of the lower value of $G_V$, our results with $G_V$=0.5 are in good agreement with the other constraints from GW170817 and PSR J0740+6620 on the $M-R$ relationship of compact stars. In Fig. \ref{mLam0p5} we find that for the moderate value of $G_V$(=0.5) the constraints on $\Lambda_{1.4}$ from both GW170817 and GW190814 are satisfied with $m_{\chi}$=100 GeV. The rest including that for the no-DM case satisfy only the one from GW190814.

\begin{figure}
\centering
\subfloat[]{\includegraphics[width=0.5\textwidth]{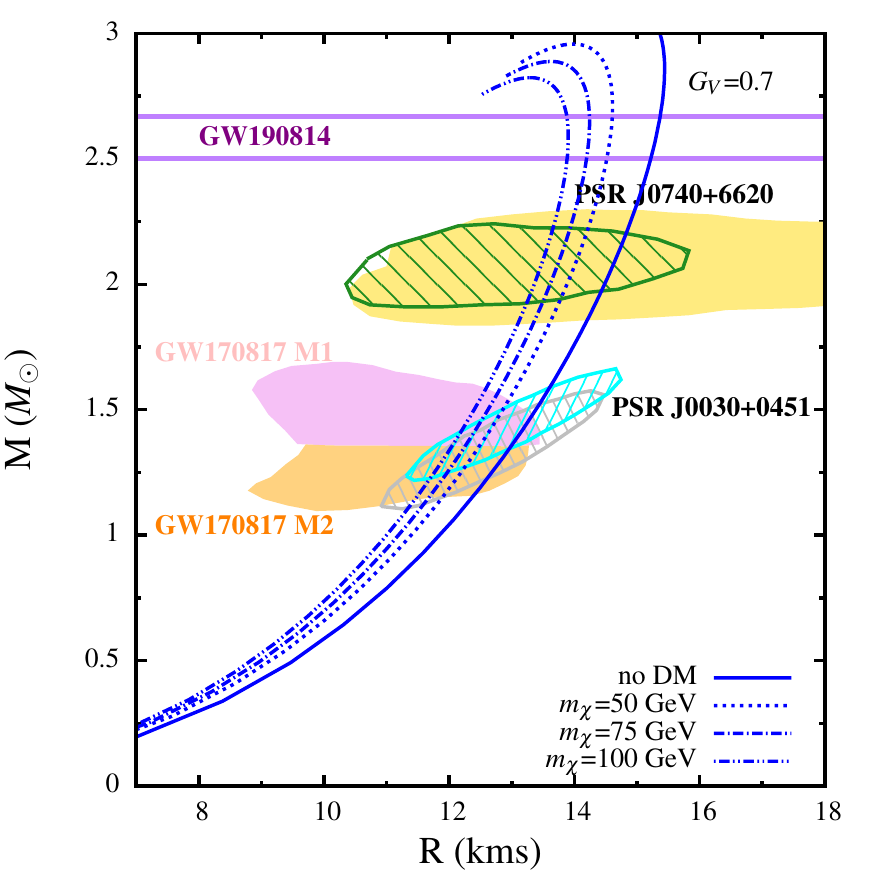}\protect\label{mr0p7}}
\vfill
\subfloat[]{\includegraphics[width=0.5\textwidth]{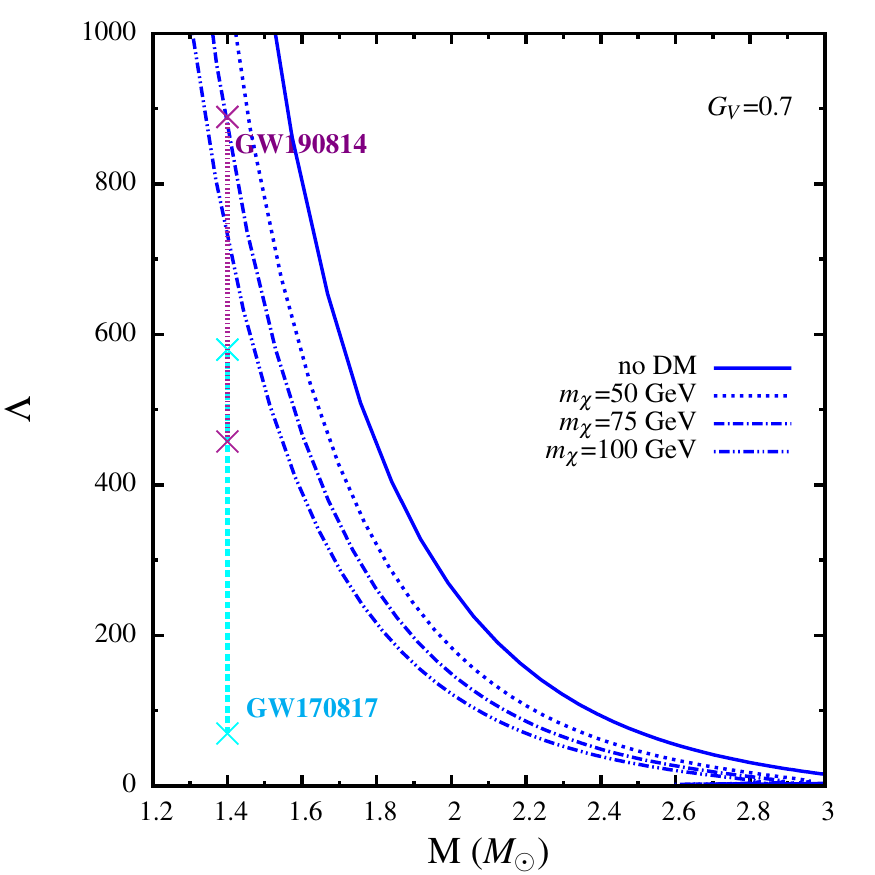}\protect\label{mLam0p7}}
\caption{\it Same as Fig. \ref{MRLam0p3} but for $G_V$=0.7.}
\label{MRLam0p7}
\end{figure}

 We finally present our results with a high value of $G_V$=0.7 in Fig. \ref{MRLam0p7}. From Fig. \ref{mr0p7} we find that $M_{max}$ offshoots the theoretical limit on the upper bound of compact stars ($>$ 3 $M_{\odot}$) in case of the pure SQS without DM. The presence of DM with all the chosen values of $m_{\chi}$ reduces the maximum mass to an acceptable range. All the constraints on the mass-radius relationship of compact stars are well fulfilled by all the DM admixed SQSs configurations. Fig. \ref{mLam0p7} suggests that for the high value of $G_V$(=0.7) the constraints on $\Lambda_{1.4}$ from GW170817 is satisfied by none of the configurations. The constraint from GW190814 is satisfied with only $m_{\chi}$=75 and 100 GeV.

 Comparing Figs. \ref{mr0p3}, \ref{mr0p5} and \ref{mr0p7}, we find that for any particular value of the $m_{\chi}$, both mass and radius of the DM admixed SQSs increase with increasing values of $G_V$. Thus we find that the NICER data for PSR J0030+0451 is better satisfied with $G_V >$ 0.3. With $G_V$ = 0.3 this constraint is marginally satisfied only for $m_{\chi}$=50 GeV. For $G_V <$ 0.3 the aforesaid constraint is not satisfied with any of the chosen values of $m_{\chi}$. We therefore chose values of $G_V >$ 0.3 in the present work. Overall, in the present work we obtain very massive SQS configurations even in the presence of DM. In most cases the maximum mass is high enough to be comparable with that of the secondary component of GW190814 merger event. Although it is at present debatable whether it is a BH or NS, the results of this work suggest that this compact object may be a DM admixed SQS. 
 
\begin{figure}
\centering
\subfloat[]{\includegraphics[width=0.5\textwidth]{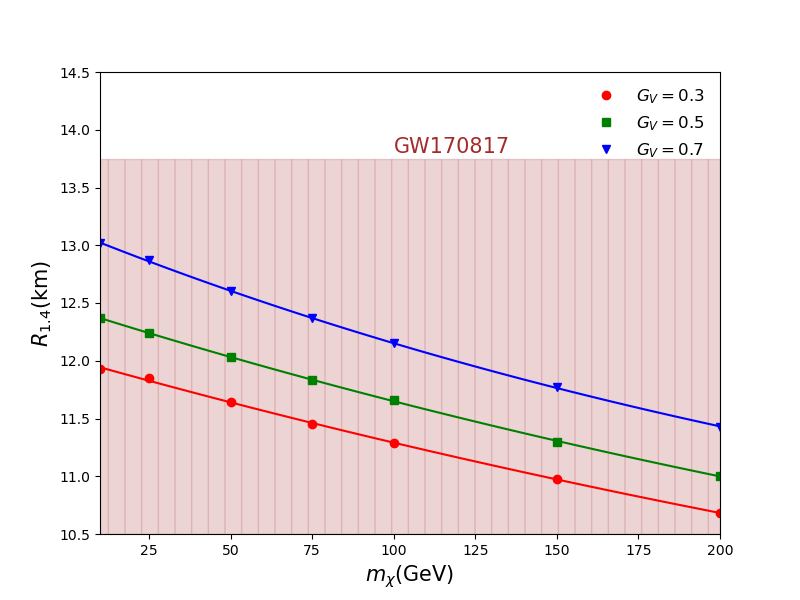}\protect\label{mchiR1p4}}
\vfill
\subfloat[]{\includegraphics[width=0.5\textwidth]{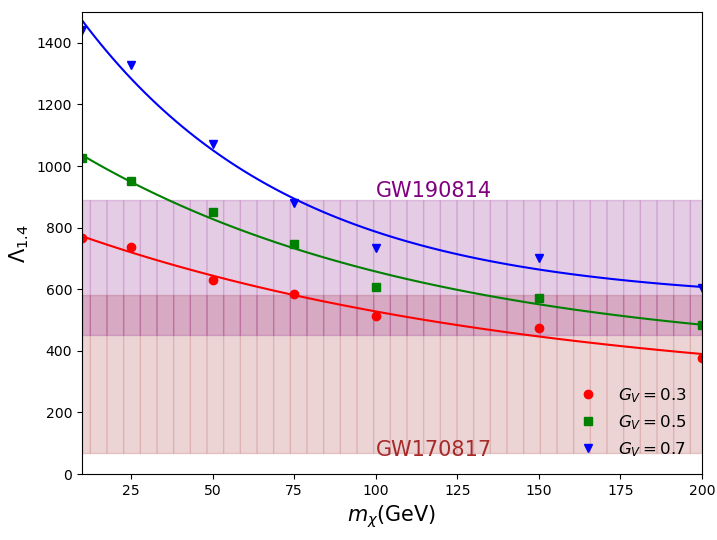}\protect\label{mchiLam1p4}}
\caption{\it Variations of (a) $R_{1.4}$ and (b) $\Lambda_{1.4}$ with respect to $m_{\chi}$ for different values of $G_V$.}
\label{mchiR1p4Lam1p4}
\end{figure}

 We next study the variation of $R_{1.4}$ and $\Lambda_{1.4}$ with respect to $m_{\chi}$. In Fig. \ref{mchiR1p4Lam1p4} we show the same. For all the values of $G_V$, both $R_{1.4}$ and $\Lambda_{1.4}$ increase with decreasing values of $m_{\chi}$. We also notice that for very high values of $m_{\chi}$ the constraint on $\Lambda_{1.4}$ from GW170817 is not well satisfied. This effect becomes more pronounced for lower values of $G_V$. Each line in Figs. \ref{mchiR1p4} and \ref{mchiLam1p4} are the fitted function for the variation of $R_{1.4}$ and $\Lambda_{1.4}$, respectively with respect to $m_{\chi}$ for the corresponding value of $G_V$. In absence of any direct relation between $R_{1.4}$ and $\Lambda_{1.4}$ with respect to $m_{\chi}$, we have obtained the specific (fitted) relationship between to $m_{\chi}$ and $R_{1.4}$ and $\Lambda_{1.4}$ individually within the scope of the model. From these fits, the particular relation between $R_{1.4}$ and $m_{\chi}$ for different $G_V$ is obtained as
  
\begin{eqnarray}
R_{1.4}=7.677 + 4.346 ~exp\Big(-\frac{m_{\chi}}{542.118}\Big)\rm{,~ for}~G_V=0.3
\label{fitR1p4_0p3}              
\end{eqnarray}

\begin{eqnarray}
R_{1.4}=8.321 + 4.138 ~exp\Big(-\frac{m_{\chi}}{459.488}\Big)\rm{,~ for}~G_V=0.5
\label{fitR1p4_0p5}              
\end{eqnarray}

and

\begin{eqnarray}
R_{1.4}=9.471 + 3.665 ~exp\Big(-\frac{m_{\chi}}{319.798}\Big)\rm{,~ for}~G_V=0.7
\label{fitR1p4_0p7}              
\end{eqnarray}

while that between $\Lambda_{1.4}$ and $m_{\chi}$ are

\begin{eqnarray}
\Lambda_{1.4}=257.08 + 553.23 ~exp\Big(-\frac{m_{\chi}}{139.95}\Big)\rm{,~ for}~G_V=0.3
\label{fitLam1p4_0p3}              
\end{eqnarray}

\begin{eqnarray}
\Lambda_{1.4}=374.21 + 726.70 ~exp\Big(-\frac{m_{\chi}}{106.13}\Big)\rm{,~ for}~G_V=0.5
\label{fitLam1p4_0p5}              
\end{eqnarray}

and

\begin{eqnarray}
\Lambda_{1.4}=558.56 + 1067.29 ~exp\Big(-\frac{m_{\chi}}{64.80}\Big)\rm{,~ for}~G_V=0.7
\label{fitLam1p4_0p7}              
\end{eqnarray}

{{It can be seen from Figs. \ref{mchiR1p4} and \ref{mchiLam1p4} that compared to the constraint on $\Lambda_{1.4}$, that on $R_{1.4}$ is better satisfied with the chosen values of $m_{\chi}$ for different $G_V$. Moreover, as mentioned earlier, that for all the chosen values of $m_{\chi}$, the NICER data for PSR J0030+0451 is better satisfied with $G_V >$ 0.3. However, there are still many uncertainties pertaining to compact star physics viz. the composition and EoS, presence of DM and its interaction with the standard model particles. Moreover, the SQM EoS is also obtained with the vbag model in the present work but there are various other well-known models to describe the SQM properties. Under such circumstances, prescribing any specific bound on the DM mass with respect to the various astrophysical constraints is beyond the scope of this work. However, within the scope of the present study, we can roughly conclude that with an average value of $m_{\chi}\sim$100 GeV the present day astrophysical constraints on compact star structural properties could be satisfied. Therefore instead of providing any specific bound on $m_{\chi}$, we have focused on obtaining the specific (fitted) relationship and dependence of $R_{1.4}$ and $\Lambda_{1.4}$ on $m_{\chi}$ for a specific value of $G_V$. Within the scope of this study the relations presented in Eqs. (\ref{fitR1p4_0p3}), (\ref{fitR1p4_0p5}) and (\ref{fitR1p4_0p7}) show the specific (fitted) relationship between $R_{1.4}$ and $m_{\chi}$ while (\ref{fitLam1p4_0p3}), (\ref{fitLam1p4_0p5}) and (\ref{fitLam1p4_0p7}) show the same between $\Lambda_{1.4}$ and $m_{\chi}$.}} 

 
 
\section{Summary and Conclusion}
\label{Conclusion}
 
 We have computed the structural properties of DM admixed SQSs. We consider the vBag model to describe the SQM with the bag constant been determined by following the stability condition of the SQM in terms of the energy density per baryon. With the notion such compact stars, being highly gravitating objects, we study the possible presence of accreted DM in SQSs thereby forming DM admixed SQSs. We introduce a feeble interaction between pure SQM and the accreted fermionic DM via a vector dark boson mediator. The masses of the DM fermion, mass of vector mediator and the coupling between them are determined in consistence with the self-interaction constraint from Bullet cluster and from present day relic abundance, respectively.
  
 Overall, we find that the presence of DM reduces both the mass and radius of the star compared to the no-DM case. We find that the presence of more massive DM particle leads to the less massive DM admixed SQSs with smaller radius. It is also seen that the various astrophysical constraints on the mass-radius relationship of the DM admixed SQSs is better satisfied with higher values of the vector coupling ($G_V$) between the quarks. With the chosen values of $m_{\chi}$ and corresponding values of $m_{\xi}$ and $y_{\xi}$, the calculated structural properties (gravitational mass, radius and tidal deformability) of the DM admixed SQS configurations satisfy all the various present day astrophysical constraints obtained from the pulsars like PSR J0740+6620 and PSR J0030+0451 and the gravitational wave data of GW170817 and GW190814.

\section*{Acknowledgments}
Work of A.G. is supported by the National Research Foundation of Korea (NRF-2019R1C1C1005073).

\section*{Data availability}

 The data underlying this article are available within the article.

\bibliographystyle{mnras}
\bibliography{ref}

\begin{thebibliography}{}
\makeatletter
\relax
\def\mn@urlcharsother{\let\do\@makeother \do\$\do\&\do\#\do\^\do\_\do\%\do\~}
\def\mn@doi{\begingroup\mn@urlcharsother \@ifnextchar [ {\mn@doi@}
  {\mn@doi@[]}}
\def\mn@doi@[#1]#2{\def\@tempa{#1}\ifx\@tempa\@empty \href
  {http://dx.doi.org/#2} {doi:#2}\else \href {http://dx.doi.org/#2} {#1}\fi
  \endgroup}
\def\mn@eprint#1#2{\mn@eprint@#1:#2::\@nil}
\def\mn@eprint@arXiv#1{\href {http://arxiv.org/abs/#1} {{\tt arXiv:#1}}}
\def\mn@eprint@dblp#1{\href {http://dblp.uni-trier.de/rec/bibtex/#1.xml}
  {dblp:#1}}
\def\mn@eprint@#1:#2:#3:#4\@nil{\def\@tempa {#1}\def\@tempb {#2}\def\@tempc
  {#3}\ifx \@tempc \@empty \let \@tempc \@tempb \let \@tempb \@tempa \fi \ifx
  \@tempb \@empty \def\@tempb {arXiv}\fi \@ifundefined
  {mn@eprint@\@tempb}{\@tempb:\@tempc}{\expandafter \expandafter \csname
  mn@eprint@\@tempb\endcsname \expandafter{\@tempc}}}

\bibitem[\protect\citeauthoryear{Aalbers et~al.}{Aalbers
  et~al.}{2022}]{LZ:2022ufs}
Aalbers J.,  et~al., 2022, arxiv, 2207.03764

\bibitem[\protect\citeauthoryear{Abbott et~al.}{Abbott
  et~al.}{2018}]{LIGOScientific:2018cki}
Abbott B.~P.,  et~al., 2018, \mn@doi [Phys. Rev. Lett.]
  {10.1103/PhysRevLett.121.161101}, 121, 161101

\bibitem[\protect\citeauthoryear{Abbott et~al.}{Abbott
  et~al.}{2020}]{LIGOScientific:2020zkf}
Abbott R.,  et~al., 2020, \mn@doi [Astrophys. J. Lett.]
  {10.3847/2041-8213/ab960f}, 896, L44

\bibitem[\protect\citeauthoryear{Aghanim et~al.}{Aghanim
  et~al.}{2020}]{Aghanim:2018eyx}
Aghanim N.,  et~al., 2020, \mn@doi [Astron. Astrophys.]
  {10.1051/0004-6361/201833910}, 641, A6

\bibitem[\protect\citeauthoryear{Agnes et~al.}{Agnes
  et~al.}{2018}]{Agnes:2018oej}
Agnes P.,  et~al., 2018, \mn@doi [Phys. Rev. Lett.]
  {10.1103/PhysRevLett.121.111303}, 121, 111303

\bibitem[\protect\citeauthoryear{Agnese et~al.}{Agnese
  et~al.}{2018}]{Agnese:2018col}
Agnese R.,  et~al., 2018, \mn@doi [Phys. Rev. Lett.]
  {10.1103/PhysRevLett.121.051301}, 121, 051301

\bibitem[\protect\citeauthoryear{Akerib et~al.}{Akerib
  et~al.}{2013}]{Akerib:2012ak}
Akerib D.~S.,  et~al., 2013, \mn@doi [Astropart. Phys.]
  {10.1016/j.astropartphys.2013.02.001}, 45, 34

\bibitem[\protect\citeauthoryear{Alford, Braby, Paris  \& Reddy}{Alford
  et~al.}{2005}]{Alford:2004pf}
Alford M.,  Braby M.,  Paris M.~W.,   Reddy S.,  2005, \mn@doi [Astrophys. J.]
  {10.1086/430902}, 629, 969

\bibitem[\protect\citeauthoryear{Aprile et~al.}{Aprile
  et~al.}{2012}]{Aprile:2012nq}
Aprile E.,  et~al., 2012, \mn@doi [Phys. Rev. Lett.]
  {10.1103/PhysRevLett.109.181301}, 109, 181301

\bibitem[\protect\citeauthoryear{Aprile et~al.}{Aprile
  et~al.}{2018}]{Aprile:2018dbl}
Aprile E.,  et~al., 2018, \mn@doi [Phys. Rev. Lett.]
  {10.1103/PhysRevLett.121.111302}, 121, 111302

\bibitem[\protect\citeauthoryear{Arcadi, Djouadi  \& Kado}{Arcadi
  et~al.}{2020}]{Arcadi:2020jqf}
Arcadi G.,  Djouadi A.,   Kado M.,  2020, \mn@doi [Phys. Lett. B]
  {10.1016/j.physletb.2020.135427}, 805, 135427

\bibitem[\protect\citeauthoryear{Aziz, Ray, Rahaman, Khlopov  \& Guha}{Aziz
  et~al.}{2019}]{Aziz:2019rgf}
Aziz A.,  Ray S.,  Rahaman F.,  Khlopov M.,   Guha B.~K.,  2019, \mn@doi [Int.
  J. Mod. Phys. D] {10.1142/S0218271819410062}, 28, 1941006

\bibitem[\protect\citeauthoryear{Bauer \& Plehn}{Bauer \&
  Plehn}{2019}]{Bauer:2017qwy}
Bauer M.,  Plehn T.,  2019, {Yet Another Introduction to Dark Matter}: {The
  Particle Physics Approach}.
 Lecture Notes in Physics Vol. 959, Springer (\mn@eprint {arXiv} {1705.01987}),
  \mn@doi{10.1007/978-3-030-16234-4}

\bibitem[\protect\citeauthoryear{Belanger, Boudjema, Pukhov  \&
  Semenov}{Belanger et~al.}{2013}]{Belanger:2013oya}
Belanger G.,  Boudjema F.,  Pukhov A.,   Semenov A.,  2013, \mn@doi [Comput.
  Phys. Commun.] {10.1016/j.cpc.2013.10.016}, 185

\bibitem[\protect\citeauthoryear{Bennett et~al.}{Bennett
  et~al.}{2013}]{WMAP:2012fli}
Bennett C.~L.,  et~al., 2013, \mn@doi [Astrophys. J. Suppl.]
  {10.1088/0067-0049/208/2/20}, 208, 20

\bibitem[\protect\citeauthoryear{Bertone, Hooper  \& Silk}{Bertone
  et~al.}{2005}]{Bertone:2004pz}
Bertone G.,  Hooper D.,   Silk J.,  2005, \mn@doi [Phys. Rept.]
  {10.1016/j.physrep.2004.08.031}, 405, 279

\bibitem[\protect\citeauthoryear{Bertoni, Nelson  \& Reddy}{Bertoni
  et~al.}{2013}]{Bertoni:2013bsa}
Bertoni B.,  Nelson A.~E.,   Reddy S.,  2013, \mn@doi [Phys. Rev. D]
  {10.1103/PhysRevD.88.123505}, 88, 123505

\bibitem[\protect\citeauthoryear{Bhat \& Paul}{Bhat \&
  Paul}{2020}]{Bhat:2019tnz}
Bhat S.~A.,  Paul A.,  2020, \mn@doi [Eur. Phys. J. C]
  {10.1140/epjc/s10052-020-8072-x}, 80, 544

\bibitem[\protect\citeauthoryear{Bodmer}{Bodmer}{1971}]{Bodmer:1971we}
Bodmer A.~R.,  1971, \mn@doi [Phys. Rev. D] {10.1103/PhysRevD.4.1601}, 4, 1601

\bibitem[\protect\citeauthoryear{Bradac et~al.,}{Bradac
  et~al.}{2006}]{Bradac:2006er}
Bradac M.,  et~al., 2006, \mn@doi [Astrophys. J.] {10.1086/508601}, 652, 937

\bibitem[\protect\citeauthoryear{Cannoni}{Cannoni}{2016}]{Cannoni:2015wba}
Cannoni M.,  2016, \mn@doi [Eur. Phys. J. C] {10.1140/epjc/s10052-016-3991-2},
  76, 137

\bibitem[\protect\citeauthoryear{Chin \& Kerman}{Chin \&
  Kerman}{1979}]{Chin:1979yb}
Chin S.~A.,  Kerman A.~K.,  1979, \mn@doi [Phys. Rev. Lett.]
  {10.1103/PhysRevLett.43.1292}, 43, 1292

\bibitem[\protect\citeauthoryear{Chodos, Jaffe, Johnson, Thorn  \&
  Weisskopf}{Chodos et~al.}{1974}]{Chodos:1974je}
Chodos A.,  Jaffe R.~L.,  Johnson K.,  Thorn C.~B.,   Weisskopf V.~F.,  1974,
  \mn@doi [Phys. Rev. D] {10.1103/PhysRevD.9.3471}, 9, 3471

\bibitem[\protect\citeauthoryear{Cierniak, Fischer, Bastian, Kl\"ahn  \&
  Salinas}{Cierniak et~al.}{2019}]{Cierniak:2019hhe}
Cierniak M.,  Fischer T.,  Bastian N.-U.,  Kl\"ahn T.,   Salinas M.,  2019,
  \mn@doi [Universe] {10.3390/universe5080186}, 5, 186

\bibitem[\protect\citeauthoryear{Crisler, Essig, Estrada, Fernandez,
  Tiffenberg, Sofo~haro, Volansky  \& Yu}{Crisler
  et~al.}{2018}]{Crisler:2018gci}
Crisler M.,  Essig R.,  Estrada J.,  Fernandez G.,  Tiffenberg J.,  Sofo~haro
  M.,  Volansky T.,   Yu T.-T.,  2018, \mn@doi [Phys. Rev. Lett.]
  {10.1103/PhysRevLett.121.061803}, 121, 061803

\bibitem[\protect\citeauthoryear{Das, Malik  \& Nayak}{Das
  et~al.}{2019}]{Das:2018frc}
Das A.,  Malik T.,   Nayak A.~C.,  2019, \mn@doi [Phys. Rev. D]
  {10.1103/PhysRevD.99.043016}, 99, 043016

\bibitem[\protect\citeauthoryear{Das, Kumar, Kumar, Kumar~Biswal, Nakatsukasa,
  Li  \& Patra}{Das et~al.}{2020}]{Das:2020vng}
Das H.~C.,  Kumar A.,  Kumar B.,  Kumar~Biswal S.,  Nakatsukasa T.,  Li A.,
  Patra S.~K.,  2020, \mn@doi [Mon. Not. Roy. Astron. Soc.]
  {10.1093/mnras/staa1435}, 495, 4893

\bibitem[\protect\citeauthoryear{Das, Kumar  \& Patra}{Das
  et~al.}{2021}]{Das:2021yny}
Das H.~C.,  Kumar A.,   Patra S.~K.,  2021, \mn@doi [Phys. Rev. D]
  {10.1103/PhysRevD.104.063028}, 104, 063028

\bibitem[\protect\citeauthoryear{Dave, Spergel, Steinhardt  \& Wandelt}{Dave
  et~al.}{2001}]{Dave:2000ar}
Dave R.,  Spergel D.~N.,  Steinhardt P.~J.,   Wandelt B.~D.,  2001, \mn@doi
  [Astrophys. J.] {10.1086/318417}, 547, 574

\bibitem[\protect\citeauthoryear{Dawson et~al.}{Dawson
  et~al.}{2012}]{Dawson:2011kf}
Dawson W.~A.,  et~al., 2012, \mn@doi [Astrophys. J. Lett.]
  {10.1088/2041-8205/747/2/L42}, 747, L42

\bibitem[\protect\citeauthoryear{Deliyergiyev, Del~Popolo, Tolos, Le~Delliou,
  Lee  \& Burgio}{Deliyergiyev et~al.}{2019}]{Deliyergiyev:2019vti}
Deliyergiyev M.,  Del~Popolo A.,  Tolos L.,  Le~Delliou M.,  Lee X.,   Burgio
  F.,  2019, \mn@doi [Phys. Rev. D] {10.1103/PhysRevD.99.063015}, 99, 063015

\bibitem[\protect\citeauthoryear{Dengler, Schaffner-Bielich  \& Tolos}{Dengler
  et~al.}{2022}]{Dengler:2021qcq}
Dengler Y.,  Schaffner-Bielich J.,   Tolos L.,  2022, \mn@doi [Phys. Rev. D]
  {10.1103/PhysRevD.105.043013}, 105, 043013

\bibitem[\protect\citeauthoryear{Ellis, H\"utsi, Kannike, Marzola, Raidal  \&
  Vaskonen}{Ellis et~al.}{2018}]{Ellis:2018bkr}
Ellis J.,  H\"utsi G.,  Kannike K.,  Marzola L.,  Raidal M.,   Vaskonen V.,
  2018, \mn@doi [Phys. Rev. D] {10.1103/PhysRevD.97.123007}, 97, 123007

\bibitem[\protect\citeauthoryear{Farhi \& Jaffe}{Farhi \&
  Jaffe}{1984}]{Farhi:1984qu}
Farhi E.,  Jaffe R.~L.,  1984, \mn@doi [Phys. Rev. D]
  {10.1103/PhysRevD.30.2379}, 30, 2379

\bibitem[\protect\citeauthoryear{Ferrer, de~la Incera  \& Paulucci}{Ferrer
  et~al.}{2015}]{Ferrer:2015vca}
Ferrer E.~J.,  de~la Incera V.,   Paulucci L.,  2015, \mn@doi [Phys. Rev. D]
  {10.1103/PhysRevD.92.043010}, 92, 043010

\bibitem[\protect\citeauthoryear{Fonseca et~al.}{Fonseca
  et~al.}{2021}]{Fonseca:2021wxt}
Fonseca E.,  et~al., 2021, \mn@doi [Astrophys. J. Lett.]
  {10.3847/2041-8213/ac03b8}, 915, L12

\bibitem[\protect\citeauthoryear{Fraga, Pisarski  \& Schaffner-Bielich}{Fraga
  et~al.}{2001}]{Fraga:2001id}
Fraga E.~S.,  Pisarski R.~D.,   Schaffner-Bielich J.,  2001, \mn@doi [Phys.
  Rev. D] {10.1103/PhysRevD.63.121702}, 63, 121702

\bibitem[\protect\citeauthoryear{Franzon, Gomes  \& Schramm}{Franzon
  et~al.}{2016}]{Franzon:2016urz}
Franzon B.,  Gomes R.~O.,   Schramm S.,  2016, \mn@doi [Mon. Not. Roy. Astron.
  Soc.] {10.1093/mnras/stw1967}, 463, 571

\bibitem[\protect\citeauthoryear{Gabrielli, Marzola, Raidal  \&
  Veerm\"ae}{Gabrielli et~al.}{2015}]{Gabrielli:2015hua}
Gabrielli E.,  Marzola L.,  Raidal M.,   Veerm\"ae H.,  2015, \mn@doi [JHEP]
  {10.1007/JHEP08(2015)150}, 08, 150

\bibitem[\protect\citeauthoryear{Gleason, Brown  \& Kain}{Gleason
  et~al.}{2022}]{Gleason:2022eeg}
Gleason T.,  Brown B.,   Kain B.,  2022, \mn@doi [Phys. Rev. D]
  {10.1103/PhysRevD.105.023010}, 105, 023010

\bibitem[\protect\citeauthoryear{Glendenning}{Glendenning}{2000}]{Glendenning:1997wn}
Glendenning N.~K.,  2000, {Compact stars: Nuclear physics, particle physics,
  and general relativity}.
Springer-Verlag, New York

\bibitem[\protect\citeauthoryear{Gondolo \& Gelmini}{Gondolo \&
  Gelmini}{1991}]{Gondolo:1990dk}
Gondolo P.,  Gelmini G.,  1991, \mn@doi [Nucl. Phys. B]
  {10.1016/0550-3213(91)90438-4}, 360, 145

\bibitem[\protect\citeauthoryear{Guha \& Sen}{Guha \& Sen}{2021}]{Guha:2021njn}
Guha A.,  Sen D.,  2021, \mn@doi [JCAP] {10.1088/1475-7516/2021/09/027}, 09,
  027

\bibitem[\protect\citeauthoryear{Guha, Dev  \& Das}{Guha
  et~al.}{2019}]{Guha:2018mli}
Guha A.,  Dev P. S.~B.,   Das P.~K.,  2019, \mn@doi [JCAP]
  {10.1088/1475-7516/2019/02/032}, 02, 032

\bibitem[\protect\citeauthoryear{Hambye \& Vanderheyden}{Hambye \&
  Vanderheyden}{2020}]{Hambye:2019tjt}
Hambye T.,  Vanderheyden L.,  2020, \mn@doi [JCAP]
  {10.1088/1475-7516/2020/05/001}, 05, 001

\bibitem[\protect\citeauthoryear{Hinderer}{Hinderer}{2008}]{Hinderer:2007mb}
Hinderer T.,  2008, \mn@doi [Astrophys. J.] {10.1086/533487}, 677, 1216

\bibitem[\protect\citeauthoryear{Hinderer, Lackey, Lang  \& Read}{Hinderer
  et~al.}{2010}]{Hinderer:2009ca}
Hinderer T.,  Lackey B.~D.,  Lang R.~N.,   Read J.~S.,  2010, \mn@doi [Phys.
  Rev. D] {10.1103/PhysRevD.81.123016}, 81, 123016

\bibitem[\protect\citeauthoryear{Jim\'enez \& Fraga}{Jim\'enez \&
  Fraga}{2022}]{Jimenez:2021nmr}
Jim\'enez J.~C.,  Fraga E.~S.,  2022, \mn@doi [Universe]
  {10.3390/universe8010034}, 8, 34

\bibitem[\protect\citeauthoryear{Kahlhoefer, Schmidt-Hoberg, Kummer  \&
  Sarkar}{Kahlhoefer et~al.}{2015}]{Kahlhoefer:2015vua}
Kahlhoefer F.,  Schmidt-Hoberg K.,  Kummer J.,   Sarkar S.,  2015, \mn@doi
  [Mon. Not. Roy. Astron. Soc.] {10.1093/mnrasl/slv088}, 452, L54

\bibitem[\protect\citeauthoryear{Karkevandi, Shakeri, Sagun  \&
  Ivanytskyi}{Karkevandi et~al.}{2022}]{Karkevandi:2021ygv}
Karkevandi D.~R.,  Shakeri S.,  Sagun V.,   Ivanytskyi O.,  2022, \mn@doi
  [Phys. Rev. D] {10.1103/PhysRevD.105.023001}, 105, 023001

\bibitem[\protect\citeauthoryear{Klahn \& Fischer}{Klahn \&
  Fischer}{2015}]{Klahn:2015mfa}
Klahn T.,  Fischer T.,  2015, \mn@doi [Astrophys. J.]
  {10.1088/0004-637X/810/2/134}, 810, 134

\bibitem[\protect\citeauthoryear{Kumar, Thapa  \& Sinha}{Kumar
  et~al.}{2022}]{Kumar:2022byc}
Kumar A.,  Thapa V.~B.,   Sinha M.,  2022, \mn@doi [Mon. Not. Roy. Astron.
  Soc.] {10.1093/mnras/stac1150}, 513, 3788

\bibitem[\protect\citeauthoryear{Leung, Chu  \& Lin}{Leung
  et~al.}{2022}]{Leung:2022wcf}
Leung K.-L.,  Chu M.-c.,   Lin L.-M.,  2022, \mn@doi [Phys. Rev. D]
  {10.1103/PhysRevD.105.123010}, 105, 123010

\bibitem[\protect\citeauthoryear{Li, Huang  \& Xu}{Li et~al.}{2012}]{Li:2012ii}
Li A.,  Huang F.,   Xu R.-X.,  2012, \mn@doi [Astropart. Phys.]
  {10.1016/j.astropartphys.2012.07.006}, 37, 70

\bibitem[\protect\citeauthoryear{Lopes \& Panotopoulos}{Lopes \&
  Panotopoulos}{2018}]{Lopes:2018oao}
Lopes I.,  Panotopoulos G.,  2018, \mn@doi [Phys. Rev. D]
  {10.1103/PhysRevD.97.024030}, 97, 024030

\bibitem[\protect\citeauthoryear{Lopes, Biesdorf  \& Menezes}{Lopes
  et~al.}{2021}]{Lopes:2020btp}
Lopes L.~L.,  Biesdorf C.,   Menezes D. e.~P.,  2021, \mn@doi [Phys. Scripta]
  {10.1088/1402-4896/abef34}, 96, 065303

\bibitem[\protect\citeauthoryear{Louren\c{c}o, Frederico  \&
  Dutra}{Louren\c{c}o et~al.}{2022a}]{Lourenco:2021dvh}
Louren\c{c}o O.,  Frederico T.,   Dutra M.,  2022a, \mn@doi [Phys. Rev. D]
  {10.1103/PhysRevD.105.023008}, 105, 023008

\bibitem[\protect\citeauthoryear{Louren\c{c}o, Lenzi, Frederico  \&
  Dutra}{Louren\c{c}o et~al.}{2022b}]{Lourenco:2022fmf}
Louren\c{c}o O.,  Lenzi C.~H.,  Frederico T.,   Dutra M.,  2022b, \mn@doi
  [Phys. Rev. D] {10.1103/PhysRevD.106.043010}, 106, 043010

\bibitem[\protect\citeauthoryear{Miao, Zhu, Li  \& Huang}{Miao
  et~al.}{2022}]{Miao:2022rqj}
Miao Z.,  Zhu Y.,  Li A.,   Huang F.,  2022, \mn@doi [Astrophys. J.]
  {10.3847/1538-4357/ac8544}, 936, 69

\bibitem[\protect\citeauthoryear{Miller et~al.}{Miller
  et~al.}{2019}]{Miller:2019cac}
Miller M.~C.,  et~al., 2019, \mn@doi [Astrophys. J. Lett.]
  {10.3847/2041-8213/ab50c5}, 887, L24

\bibitem[\protect\citeauthoryear{Miller et~al.}{Miller
  et~al.}{2021}]{Miller:2021qha}
Miller M.~C.,  et~al., 2021, \mn@doi [Astrophys. J. Lett.]
  {10.3847/2041-8213/ac089b}, 918, L28

\bibitem[\protect\citeauthoryear{Mukhopadhyay \&
  Schaffner-Bielich}{Mukhopadhyay \&
  Schaffner-Bielich}{2016}]{Mukhopadhyay:2015xhs}
Mukhopadhyay P.,  Schaffner-Bielich J.,  2016, \mn@doi [Phys. Rev. D]
  {10.1103/PhysRevD.93.083009}, 93, 083009

\bibitem[\protect\citeauthoryear{Mukhopadhyay, Atta, Imam, Basu  \&
  Samanta}{Mukhopadhyay et~al.}{2017}]{Mukhopadhyay:2016dsg}
Mukhopadhyay S.,  Atta D.,  Imam K.,  Basu D.~N.,   Samanta C.,  2017, \mn@doi
  [Eur. Phys. J. C] {10.1140/epjc/s10052-017-5006-3}, 77, 440

\bibitem[\protect\citeauthoryear{Nandi \& Char}{Nandi \&
  Char}{2018}]{Nandi:2017rhy}
Nandi R.,  Char P.,  2018, \mn@doi [Astrophys. J.] {10.3847/1538-4357/aab78c},
  857, 12

\bibitem[\protect\citeauthoryear{Nandi \& Pal}{Nandi \&
  Pal}{2021}]{Nandi:2020luz}
Nandi R.,  Pal S.,  2021, \mn@doi [Eur. Phys. J. ST]
  {10.1140/epjs/s11734-021-00004-4}, 230, 551

\bibitem[\protect\citeauthoryear{Nelson, Reddy  \& Zhou}{Nelson
  et~al.}{2019}]{Nelson:2018xtr}
Nelson A.,  Reddy S.,   Zhou D.,  2019, \mn@doi [JCAP]
  {10.1088/1475-7516/2019/07/012}, 07, 012

\bibitem[\protect\citeauthoryear{Olinto}{Olinto}{1987}]{Olinto:1986je}
Olinto A.~V.,  1987, \mn@doi [Phys. Lett. B] {10.1016/0370-2693(87)91144-0},
  192, 71

\bibitem[\protect\citeauthoryear{Oppenheimer \& Volkoff}{Oppenheimer \&
  Volkoff}{1939}]{Oppenheimer:1939ne}
Oppenheimer J.~R.,  Volkoff G.~M.,  1939, \mn@doi [Phys. Rev.]
  {10.1103/PhysRev.55.374}, 55, 374

\bibitem[\protect\citeauthoryear{Panotopoulos \& Lopes}{Panotopoulos \&
  Lopes}{2017a}]{Panotopoulos:2017pgv}
Panotopoulos G.,  Lopes I.,  2017a, \mn@doi [Phys. Rev. D]
  {10.1103/PhysRevD.96.023002}, 96, 023002

\bibitem[\protect\citeauthoryear{Panotopoulos \& Lopes}{Panotopoulos \&
  Lopes}{2017b}]{Panotopoulos:2017idn}
Panotopoulos G.,  Lopes I.,  2017b, \mn@doi [Phys. Rev. D]
  {10.1103/PhysRevD.96.083004}, 96, 083004

\bibitem[\protect\citeauthoryear{Panotopoulos \& Lopes}{Panotopoulos \&
  Lopes}{2017c}]{Panotopoulos:2017eig}
Panotopoulos G.,  Lopes I.,  2017c, \mn@doi [Phys. Rev. D]
  {10.1103/PhysRevD.96.083013}, 96, 083013

\bibitem[\protect\citeauthoryear{Panotopoulos \& Lopes}{Panotopoulos \&
  Lopes}{2018a}]{Panotopoulos:2018joc}
Panotopoulos G.,  Lopes I.,  2018a, \mn@doi [Int. J. Mod. Phys. D]
  {10.1142/S0218271818500931}, 27, 1850093

\bibitem[\protect\citeauthoryear{Panotopoulos \& Lopes}{Panotopoulos \&
  Lopes}{2018b}]{Panotopoulos:2018ipq}
Panotopoulos G.,  Lopes I.,  2018b, \mn@doi [Phys. Rev. D]
  {10.1103/PhysRevD.98.083001}, 98, 083001

\bibitem[\protect\citeauthoryear{Quddus, Panotopoulos, Kumar, Ahmad  \&
  Patra}{Quddus et~al.}{2020}]{Quddus:2019ghy}
Quddus A.,  Panotopoulos G.,  Kumar B.,  Ahmad S.,   Patra S.~K.,  2020,
  \mn@doi [J. Phys. G] {10.1088/1361-6471/ab9d36}, 47, 095202

\bibitem[\protect\citeauthoryear{Randall, Markevitch, Clowe, Gonzalez  \&
  Bradac}{Randall et~al.}{2008}]{Randall:2007ph}
Randall S.~W.,  Markevitch M.,  Clowe D.,  Gonzalez A.~H.,   Bradac M.,  2008,
  \mn@doi [Astrophys. J.] {10.1086/587859}, 679, 1173

\bibitem[\protect\citeauthoryear{Rezaei}{Rezaei}{2017}]{Rezaei:2016zje}
Rezaei Z.,  2017, \mn@doi [Astrophys. J.] {10.1088/1361-6528/aa5273}, 835, 33

\bibitem[\protect\citeauthoryear{Riley et~al.}{Riley
  et~al.}{2019}]{Riley:2019yda}
Riley T.~E.,  et~al., 2019, \mn@doi [Astrophys. J. Lett.]
  {10.3847/2041-8213/ab481c}, 887, L21

\bibitem[\protect\citeauthoryear{Riley et~al.}{Riley
  et~al.}{2021}]{Riley:2021pdl}
Riley T.~E.,  et~al., 2021, \mn@doi [Astrophys. J. Lett.]
  {10.3847/2041-8213/ac0a81}, 918, L27

\bibitem[\protect\citeauthoryear{Robertson, Massey  \& Eke}{Robertson
  et~al.}{2017}]{Robertson:2016xjh}
Robertson A.,  Massey R.,   Eke V.,  2017, \mn@doi [Mon. Not. Roy. Astron.
  Soc.] {10.1093/mnras/stw2670}, 465, 569

\bibitem[\protect\citeauthoryear{Sen}{Sen}{2021}]{Sen:2020edi}
Sen D.,  2021, \mn@doi [J. Phys. G] {10.1088/1361-6471/abcb9e}, 48, 025201

\bibitem[\protect\citeauthoryear{Sen \& Guha}{Sen \& Guha}{2021}]{Sen:2021wev}
Sen D.,  Guha A.,  2021, \mn@doi [Mon. Not. Roy. Astron. Soc.]
  {10.1093/mnras/stab1056}, 504, 3354

\bibitem[\protect\citeauthoryear{Tanabashi et~al.}{Tanabashi
  et~al.}{2018}]{ParticleDataGroup:2018ovx}
Tanabashi M.,  et~al., 2018, \mn@doi [Phys. Rev. D]
  {10.1103/PhysRevD.98.030001}, 98, 030001

\bibitem[\protect\citeauthoryear{Tolman}{Tolman}{1939}]{Tolman:1939jz}
Tolman R.~C.,  1939, \mn@doi [Phys. Rev.] {10.1103/PhysRev.55.364}, 55, 364

\bibitem[\protect\citeauthoryear{Tolos \& Schaffner-Bielich}{Tolos \&
  Schaffner-Bielich}{2015}]{Tolos:2015qra}
Tolos L.,  Schaffner-Bielich J.,  2015, \mn@doi [Phys. Rev. D]
  {10.1103/PhysRevD.92.123002}, 92, 123002

\bibitem[\protect\citeauthoryear{Torres \& Menezes}{Torres \&
  Menezes}{2013}]{Torres:2012xv}
Torres J.~R.,  Menezes D.~P.,  2013, \mn@doi [EPL]
  {10.1209/0295-5075/101/42003}, 101, 42003

\bibitem[\protect\citeauthoryear{Tulin \& Yu}{Tulin \&
  Yu}{2018}]{Tulin:2017ara}
Tulin S.,  Yu H.-B.,  2018, \mn@doi [Phys. Rept.]
  {10.1016/j.physrep.2017.11.004}, 730, 1

\bibitem[\protect\citeauthoryear{Tulin, Yu  \& Zurek}{Tulin
  et~al.}{2013}]{Tulin:2013teo}
Tulin S.,  Yu H.-B.,   Zurek K.~M.,  2013, \mn@doi [Phys. Rev. D]
  {10.1103/PhysRevD.87.115007}, 87, 115007

\bibitem[\protect\citeauthoryear{Vogelsberger, Zavala  \& Loeb}{Vogelsberger
  et~al.}{2012}]{Vogelsberger:2012ku}
Vogelsberger M.,  Zavala J.,   Loeb A.,  2012, \mn@doi [Mon. Not. Roy. Astron.
  Soc.] {10.1111/j.1365-2966.2012.21182.x}, 423, 3740

\bibitem[\protect\citeauthoryear{Wang et~al.}{Wang et~al.}{2020}]{Wang:2020coa}
Wang Q.,  et~al., 2020, \mn@doi [Chin. Phys. C] {10.1088/1674-1137/abb658}, 44,
  125001

\bibitem[\protect\citeauthoryear{Wei, Irving, Kl\"ahn  \& Jaikumar}{Wei
  et~al.}{2019}]{Wei:2018mxy}
Wei W.,  Irving B.,  Kl\"ahn T.,   Jaikumar P.,  2019, \mn@doi [Astrophys. J.]
  {10.3847/1538-4357/ab53ea}, 887, 151

\bibitem[\protect\citeauthoryear{Weissenborn, Sagert, Pagliara, Hempel  \&
  Schaffner-Bielich}{Weissenborn et~al.}{2011}]{Weissenborn:2011qu}
Weissenborn S.,  Sagert I.,  Pagliara G.,  Hempel M.,   Schaffner-Bielich J.,
  2011, \mn@doi [Astrophys. J. Lett.] {10.1088/2041-8205/740/1/L14}, 740, L14

\bibitem[\protect\citeauthoryear{Witten}{Witten}{1984}]{Witten:1984rs}
Witten E.,  1984, \mn@doi [Phys. Rev. D] {10.1103/PhysRevD.30.272}, 30, 272

\bibitem[\protect\citeauthoryear{Yang, PI, Zheng  \& Weber}{Yang
  et~al.}{2020}]{Yang:2019rxn}
Yang S.-H.,  PI C.-M.,  Zheng X.-P.,   Weber F.,  2020, \mn@doi [Astrophys. J.]
  {10.3847/1538-4357/abb365}, 902, 32

\bibitem[\protect\citeauthoryear{Zheng \& Chen}{Zheng \&
  Chen}{2016}]{Zheng:2016ygg}
Zheng H.,  Chen L.-W.,  2016, \mn@doi [Astrophys. J.]
  {10.3847/0004-637X/831/2/127}, 831, 127

\bibitem[\protect\citeauthoryear{Zhou, Zhou  \& Li}{Zhou
  et~al.}{2018}]{Zhou:2017pha}
Zhou E.-P.,  Zhou X.,   Li A.,  2018, \mn@doi [Phys. Rev. D]
  {10.1103/PhysRevD.97.083015}, 97, 083015

\makeatother
\end{thebibliography}

\end{document}